# Carotid Plaque Segmentation in Ultrasound Images Using a Mask R-CNN

Maxwell J. Kiernan[a*], Rashid Al Mukaddim[a], Carol C. Mitchell[b], Jenna Maybock[c], Stephanie M. Wilbrand[c], Robert J. Dempsey[c], Tomy Varghese[a]

[a]Department of Medical Physics, University of Wisconsin School of Medicine and Public health (UW-SMPH), 1111 Highland Ave #1005, Madison, WI 53705, United States
[b]Department of Medicine, UW-SMPH, 5158 Medical Foundation Centennial Building, 1685 Highland Ave, Madison, WI 53792, United States
[c]Department of Neurological Surgery, UW-SMPH, 600 Highland Avenue, K4/8 CSC, Box 8660, Madison, WI 53792, United States

## Abstract

Background: Ultrasound imaging plays a pivotal role in diagnosing carotid atherosclerosis, a significant precursor to cardiovascular and cerebrovascular diseases and events. This non-invasive modality provides real-time, high-resolution images, allowing clinicians to assess atherosclerotic plaques in the carotid arteries without invasive procedures. Early detection using ultrasound aids in timely interventions, reducing the risk of adverse cardiovascular events.

Purpose: In this study, we present the refinement of a Mask R-CNN model initially designed for carotid lumen detection to automatically generate bounding boxes (BB) enclosing atherosclerotic plaque for segmentation to assist in our ultrasound elastography workflow.

Methods: We utilize a PyTorch torchvision implementation of the Mask R-CNN for carotid plaque detection and BB placement. Our dataset consists of 118 severe stenotic carotid plaques from presenting patients, clinically indicated for a carotid endarterectomy. Due to the variability of plaque presentation in the dataset, a multitude of different R-CNN models were observed to have varying results based on the allowed number of prediction regions. An overview analysis looking at shared predictions from these models showed a slight improvement compared to the individual model results

Results: Evaluation metrics such as Dice similarity coefficient and intersection over Union are employed. The model trained with 5 maximum BB prediction regions and tested with 2 maximum BB prediction regions produced the highest individual accuracy with a Dice score of 0.74 and intersection over union of 0.61. A filtered combined analysis of all the models demonstrated a slight increase in performance with scores of 0.76 and 0.61 respectively.

Conclusion: Due to the significant variation in plaque presentation and types amongst presenting patients, the accuracy of the Plaque Mask R-CNN network would benefit from the incorporation of additional patient datasets to incorporate increased variation into the training dataset.

# Introduction

Ultrasound imaging has emerged as a crucial diagnostic tool in detecting and assessing carotid atherosclerosis, a common and significant precursor to cardiovascular and cerebrovascular diseases[1]. Its non-invasive nature and ability to provide real-time, high-resolution visualization of the carotid arteries[2] allow clinicians to identify plaque formation, stenosis, and vessel wall abnormalities without invasive procedures. This capability enables early diagnosis and risk stratification[3], which are critical for timely intervention and the prevention of adverse events such as ischemic stroke and vascular cognitive decline.

Beyond traditional structural imaging, ultrasound elastography has become a powerful modality for evaluating plaque stability.[3-6] This technique quantifies tissue strain and deformation, which are associated with plaque composition. Softer, echolucent plaques tend to exhibit higher strain and are more prone to rupture,[6-11] whereas stiffer plaques tend to be more stable. Our prior work and other studies have shown significant correlations between increased plaque strain and cognitive impairment,[7, 12] as well as changes in brain white matter hyperintensity in magnetic resonance imaging (MRI).[12] These findings suggest that non-invasive strain imaging can help identify high-risk, rupture-prone plaques and support vascular cognitive health assessment.

To perform elastography, we use a Lagrangian Carotid Strain Imaging (LCSI) protocol based on radiofrequency (RF) data, originally introduced by McCormick et al.[13] and later accelerated on GPUs by Meshram et al.[14] This method calculates strain indices within manually segmented plaques using B-mode images reconstructed from RF data. However, this reliance on manual segmentation by trained sonographers presents practical challenges. Manual annotation is time-consuming, and the requirement for expert involvement limits the scalability of our workflow to larger patient populations. In severe stenosis cases complex plaque morphologies and signal artifacts further increase segmentation difficulty. To address this limitation, the method proposed in this paper aims to replicate the segmentation of B-mode images currently performed by a trained sonographer.

Various approaches to resolving this issue have been reported. Manual image processing techniques such as active contours[15-18] and edge detection[19, 20] have shown inconsistent results and require user interaction. Both semi-automatic and automatic convolutional neural networks (CNNs) have been utilized for segmentation[21, 22]. Semi-automatic methods require manual input, such as an ROI or spline[23], and have produced favorable results due to their limited scope. However, they still require advanced knowledge to locate the proper region for segmentation[24]. Automated methods are less dependent on user inputs but have shown limitations in accuracy compared to ground truth models and require well-trained models.

In their 2021 study, Zhou et al.[21] employed a U-Net structure for segmenting carotid plaques in 3D ultrasound images, utilizing datasets from two different sources. Their study, however, excluded patients with >70% stenosis, specifically evaluated in our study, who form the more challenging group with complex plaque, attenuation, and other artifacts. Results from two independent U-Net structures demonstrated an increased level of agreement with manual segmentations, with Pearson's correlation coefficients of 0.989 (p < 0.0001) and 0.987 (p <

0.0001). Despite the apparent automation of the method, it is essential to note that the dataset employed in this study comprised pre-cropped images focusing solely on the desired plaque region. More recently, Zhou et al. (2023) [22] presented an image reconstruction-based self-supervised learning algorithm (IR-SSL) with U-Net and U-Net++, demonstrating improved performance and achieving Dice-similarity-coefficients (DSC) of 80.14–88.84%. While methods of this nature are often considered fully automatic, they still necessitate additional manual input in the form of cropping the original B-mode image. In specific scenarios, particularly when working with B-mode images, employing an automated system to crop images and retain only the ultrasound Region of Interest (ROI) may be acceptable[21, 25]. These nuances in fully automatic approaches are highlighted in our prior work conducted by Meshram et al. in 2020[24]. This work revealed that a semi-automatic approach, incorporating a user-defined bounding box to limit the search area, yielded more favorable segmentation results compared to an unrestricted fully automatic approach.

Our previous work[24] demonstrated that localizing plaque using a bounding box prior to segmentation significantly improves segmentation accuracy compared to direct segmentation from the full B-mode image. Building on this concept, the purpose of the current study was to modify and refine our existing Mask R-CNN model—originally developed for carotid lumen segmentation—to produce accurate segmentations of carotid atherosclerotic plaque automatically. Given the increased variability in plaque appearance compared to the vessel lumen, we aim to assess whether the Mask R-CNN model can achieve comparable prediction accuracy when trained specifically on plaque. Accurate plaque segmentation is critical for initializing regions of interest for downstream vessel wall and plaque analysis, ultimately resulting in more precise binary masks at end diastole.

There are three component of this study. First, we apply a Mask R-CNN model to segment plaque using B-mode images reconstructed from raw RF data—a data source that more closely reflects our elastography workflow but is rarely used in segmentation literature. Second, we systematically evaluate model performance across varying maximum bounding box constraints during training and testing, a practical but underexplored factor in regional plaque detection. Third, we propose a novel model agreement strategy that identifies consistently predicted plaque regions across multiple trained models, independent of ground truth, to improve robustness and reduce false positives. These contributions aim to support future integration of automated plaque segmentation into strain-based risk assessment protocols.

## Materials and Methods
### Human Subjects and Data
Dataset consists of 118 severe stenotic carotid plaques from presenting patients, clinically indicated for a carotid endarterectomy at University of Wisconsin-Madison as documented in our previous study[26]. Data collection utilized a Siemens S2000 ultrasound system equipped with an 18L6 linear array transducer. Ultrasound RF and B-mode data loops were obtained through freehand acquisitions employing a relatively stationary transducer. Imaging focused on three spatial locations/views—common carotid artery, carotid bifurcation, and internal carotid artery— at a depth of 4 cm on both sides for each patient. All images have a common width of 456 pixels but vary with the imaging depth used. The standard 4 cm imaging depth corresponds to a depth of 692 pixels. This process resulted in a total of 350 views displaying plaque. Subsequent manual segmentation by a sonographer identified plaque in two to three end-diastolic frames per view and is utilized as the ground truth. The segmented regions were used to derive bounding box required for training the detection model. This yielded a dataset comprising 885 individual plaque segmentations and lumen-detected frames.

### Network Structure

Utilizing a PyTorch torchvision implementation of the Mask R-CNN as described in a previous study[26] we seek to modify the existing carotid vessel detection model to achieve atherosclerotic plaque detection by first generating the bounding boxes that enclose these plaque regions, respectively. The model produces predictions from both region-based bounding box detection as well as semantic segmentations produced within the bounding boxes. For this investigation, we primarily focus on the bounding box performance for potential use with the semiautomatic plaque segmentation model described previously by Meshram et al.[24].

Previous documentation of the network determined that a model trained on solely B-mode data derived from raw RF data for the dataset produced the best segmentation results for the detection of the carotid lumen in the reshaped images. This is compared to a three-channel training model consisting of the RF-derived B-mode alongside MimickNet (MimNet) B-mode[27] and short time Spatiotemporal Clutter Filtered Power Images obtained using singular value decomposition (SVD). However, in initial investigations for using the model for carotid plaque detection, this variation in prediction accuracy is not observed in the models trained for plaque detection when using the raw (not resampled) images. The images used in this study utilize the same methods for acquisition as described in documentation for the lumen segmentation Mask R-CNN[26].

Apart from minor changes, the same training protocol from the lumen detection model is utilized. Of the 891 acquired views, 90% (771 images) were used for training and validation while the remaining 10% (117 images) were used in testing. Likewise, the training and validation sets 90% were used for training (694 images) and 10% (77 images) were used for validation. Apart from moving an image from the training dataset to the test dataset, the datasets used in this investigation are randomized so the datasets are not identical to those used in training the Lumen detection R-CNN.

This study investigates the effects of limiting the maximum number of allowed bounding box predictions on Mask R-CNN model accuracy. Different models were trained with bounding box limits ranging from 1 to 5, and each model was tested both with the same limit used during training and with a limit of 2 bounding boxes, resulting in a total of 9 distinct model configurations. The inclusion of a 2-box testing condition for all models reflects the observation that most images contain at most two plaque segments. By varying the number of bounding boxes during training and testing, we also explore how mismatches between these settings affect performance. For instance, when the model is tested with more boxes than it was trained on ($y > x$), it may output additional regions that are less optimized, potentially introducing more false positives. Conversely, testing with fewer boxes than the model was trained for ($y < x$) may suppress relevant predictions and limit the model's effectiveness. To further assess agreement across models, we identify the most consistent bounding box predictions independent of the ground truth by selecting predicted regions that appear in at least 5 out of the 9 model outputs. This majority-vote strategy highlights which regions are robustly identified across different training and testing constraints.

Both the Plaque Mask R-CNN and the Lumen Mask R-CNN utilized the ResNet50 backbone feature extraction layer and trained on a NVIDIA A40 GPU (NVIDIA Corporation, Santa Clara, CA, USA) with an AMD EPYC 7H12 CPU (Advanced Micro Devices, Inc., Santa Clara, CA, USA). The initial learning rate is unchanged (0.0025). While the learning rate reductions were kept at a 10% reduction after 16 and 22 epochs, respectively. For the Plaque Mask R-CNN, each layer of the networks is trained for 41 epochs, then the final layer of the models is trained for an additional 10 epochs for fine tuning. An example training and validation loss plots are shown in figure 1.

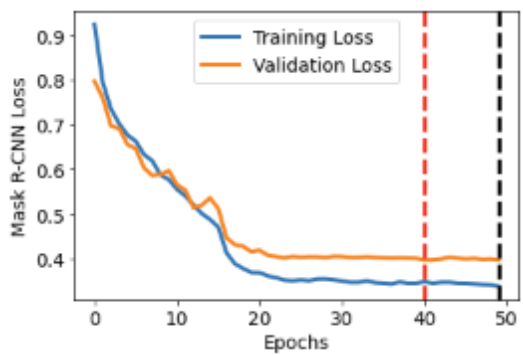

**Figure 1.** Training and Validation loss plots for the model trained with a max bounding box limit of 3. Black and red dotted lines represent the epoch in which the Training and Validation Loss respectfully, are at a minimum.

The Dice similarity coefficient (DSC), intersection over Union (IoU), average precision (AP) and average recall (AR) were used to evaluate the different models. The equations for the parameters used are shown below.

$$DSC = \frac{2TP}{2TP+FP+FN}$$ Equation 1

$$IoU = \frac{area\ of\ overlap}{area\ of\ union} = \frac{|A \cap B|}{|A \cup B|}$$ Equation 2

$$Precision = \frac{TP}{TP+FP}$$ Equation 3

$$Recall = \frac{TP}{TP+FN}$$ Equation 4

TP indicates a true positive result, where the predicted bounding box/segmentation region agrees with the ground truth. FP indicates a false positive result, where the model predicts a result that does not agree with the ground truth. FN indicates a false negative result, where the model does not predict a result, but a result is expected in the ground truth. A and B in the IoU equation represent the predicted and ground truth regions respectfully.

Statistical analysis compared the DSC and IoU of the results of each run to the results found in the most accurate lumen segmentation model using z-testing. If the model is to be improved over the lumen segmentation model, the bounding box results should produce a DSC of greater than 0.89 and IoU of 0.81, while the segmentation results should have an DSC of greater than 0.85 and IoU of 0.75.

# Results

Figure 2 contains a few representative examples for 3 selected models, a model trained with a maximum of three bounding boxes and tested on a maximum of two bounding boxes (A-C) and models which were trained with a maximum of five bounding boxes and tested with a maximum of 2 (D-F) and 5 bounding boxes (G-I) respectively. For each model there are three cases labeled as columns I, II, and III. Each instance contains a bounding box prediction and a segmentation prediction. The bounding box prediction contains the RF-derived B-mode image with the baseline segmentations for the carotid lumen and any carotid plaque present. The ground truth bounding boxes derived from the ground truth plaque segmentations are depicted with a blue rectangular outline, while the predicted bounding box is depicted as a red outline. The segmentation results are shown with the ground truth manual segmentations shown in dark yellow, and the predicted segmentation maps are overlaid within the predicted regions using a scale from green to yellow, where brighter yellow denotes more confidence in the result.

Figure 3 presents bounding box predictions for all models tested. For each image, corresponding to the respective columns *I, II,* and *III* in Figure 2, the ground truth bounding box area, raw predicted bounding box area, and the filtered results are shown. The first image is simply the ground truth bounding box area for reference. The second image is a raw composite of all nine segmentation models tested in this study. Brighter regions indicate areas of high agreement between models. The final image contains the final 'reconstructed' result by only incorporating regions where at least five of the models agree upon an area.

Table 1 presents bounding box results for the nine training and testing schemes. For each model both mean DSC and mean IoU are shown along with their standard deviations. The final two rows account for the raw and filtered combined bounding box results shown in Figure 3. The raw predictions compare the ground truth with the raw unfiltered results. Likewise, the filtered predictions compare the ground truth with the filtered results which only accounts for regions where at least five of the models agree upon an area. Though not the focus of this particular study, the same evaluation metric used in Table 1 is utilized in Table 2 for evaluation of segmentation result for each of the models. For both tables, higher values of DSC, IoU, AP, and AR indicate better results.

Table 3 presents the comparison between three-channel and single-channel input data. The mean IoU and DSC results for the same models shown in Tables 1 and 2. Statistical significance for each of the listed models indicated that each of the models produced statistically lower accuracy when compared to the accuracy of the results lumen segmentation model as all had a p value of <0.001. Performance of the best plaque models are compared to the performance of the lumen model in Table 4.

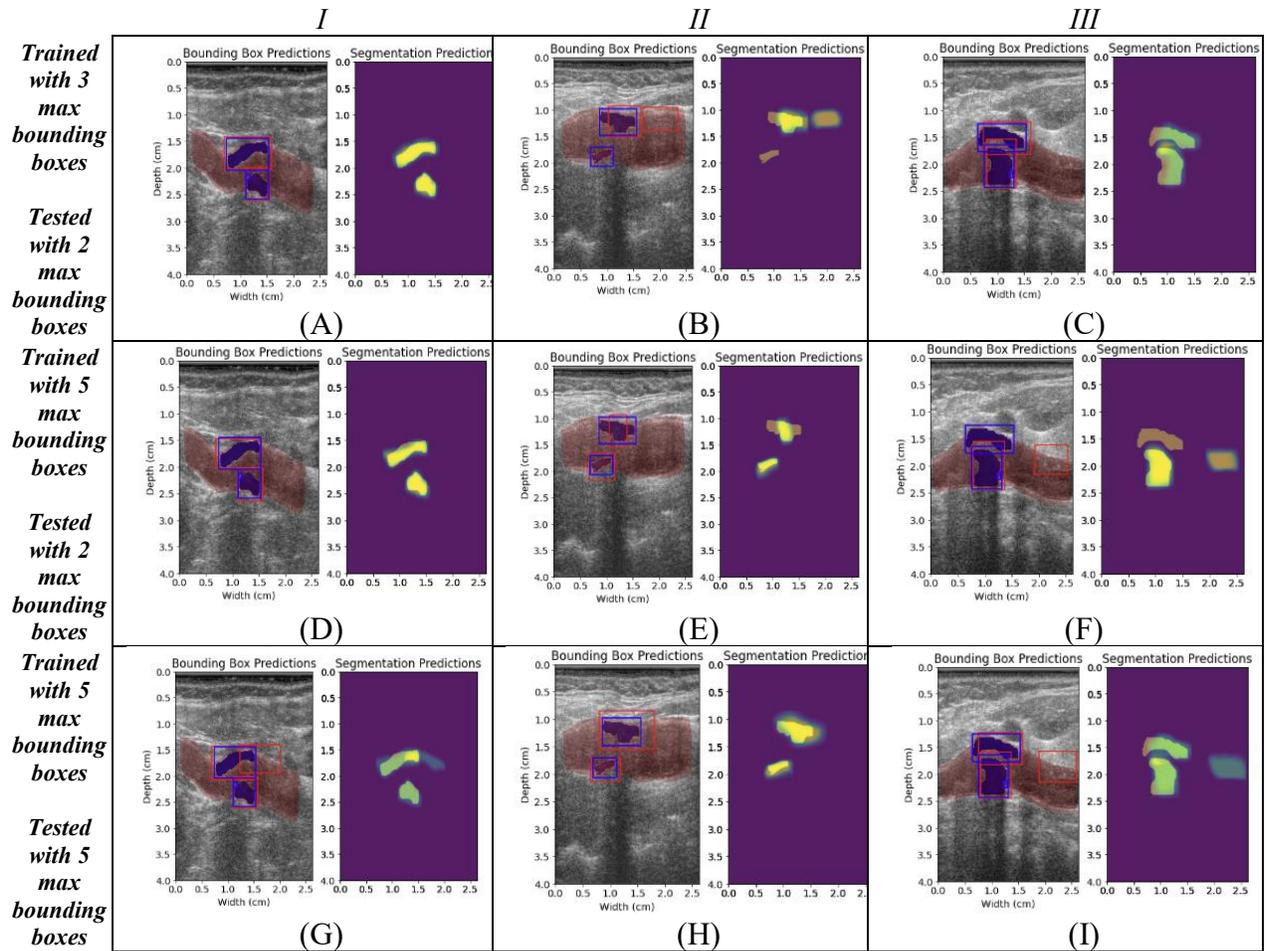

**Figure 2:** Example bounding box and segmentation results for three select cases for three models. A model trained on 3 boxes and tested on 2 boxes (A-C), 5 boxes and 2 boxes (D-F), and 5 boxes and 5 boxes (G-I). Images in columns *I, II,* and *III* correspond to the same test case.

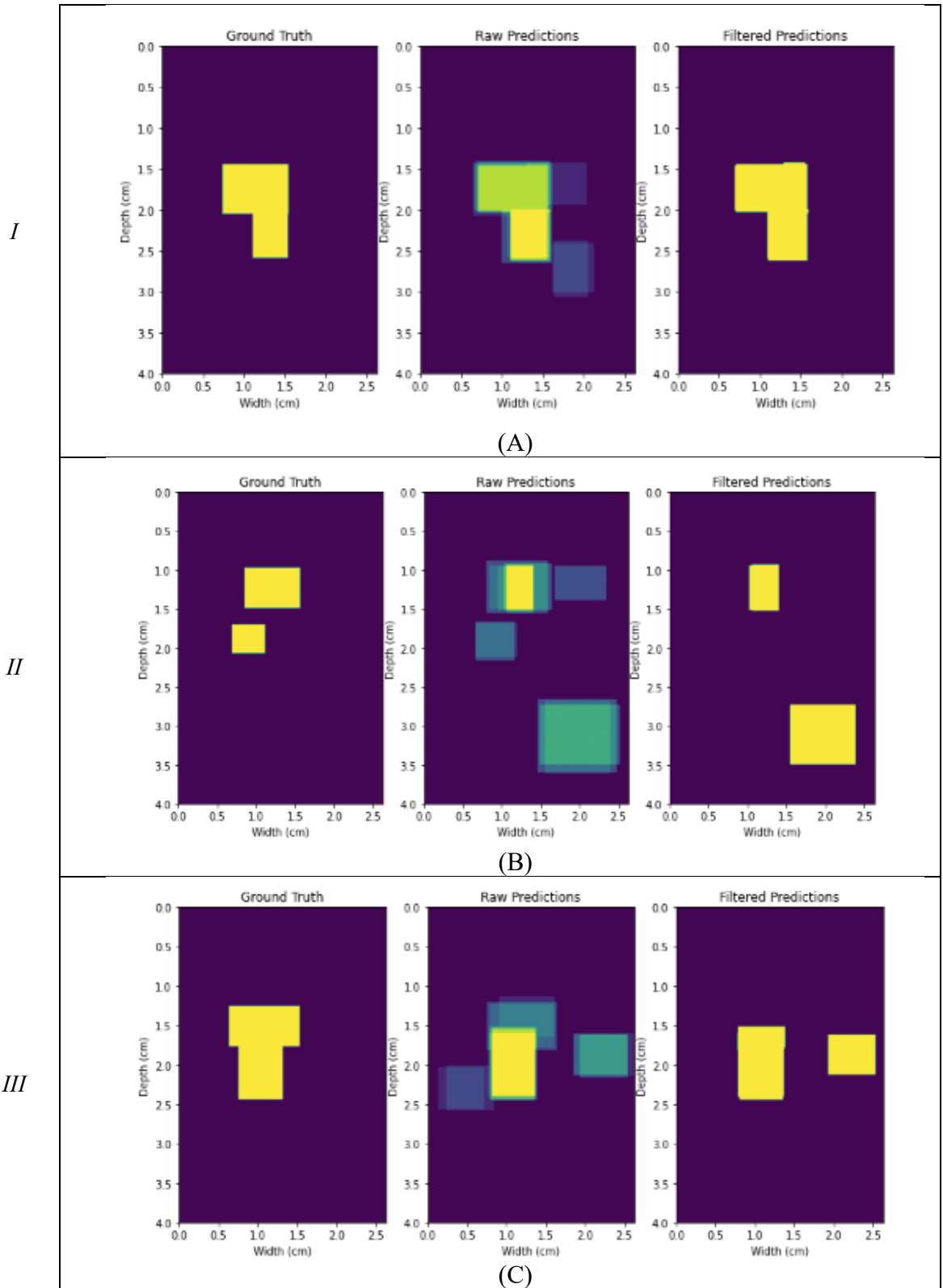

**Figure 3.** Bounding box analysis images. Ground truth, raw predictions, and filtered predictions for the corresponding cases in Figure 2. (A) image *I,* (B) image *II,* (C) image *III*

| Max Training Boxes | Max Testing Boxes | DSC | IoU | AP IoU=0.50:0.95 | AP IoU=0.50 | AP IoU=0.75 | AR IoU=0.50:0.95 |
|---|---|---|---|---|---|---|---|
| *1* | *1* | 0.62±0.22 | 0.48±0.22 | 0.107 | 0.215 | 0.074 | 0.128 |
| *2* | *2* | 0.72±0.20 | 0.59±0.22 | 0.165 | 0.331 | 0.156 | 0.202 |
| *3* | *3* | 0.69±0.18 | 0.56±0.20 | 0.171 | 0.338 | 0.170 | 0.216 |
| *4* | *4* | 0.65±0.17 | 0.51±0.20 | 0.165 | 0.335 | 0.146 | 0.212 |
| *5* | *5* | 0.67±0.19 | 0.53±0.19 | 0.165 | 0.345 | 0.144 | 0.217 |
| *1* | *2* | 0.72±0.20 | 0.60±0.22 | 0.163 | 0.349 | 0.115 | 0.201 |
| *3* | *2* | 0.72±0.19 | 0.60±0.22 | 0.165 | 0.324 | 0.171 | 0.207 |
| *4* | *2* | 0.71±0.20 | 0.59±0.22 | 0.160 | 0.318 | 0.143 | 0.203 |
| *5* | *2* | 0.74±0.17 | 0.61±0.20 | 0.159 | 0.334 | 0.140 | 0.205 |
| *Raw Combined* | | 0.69±0.18 | 0.58±0.20 | | | | |
| *Filtered Predictions* | | 0.76±0.17 | 0.61±0.21 | | | | |

**Table 1.** Network performance statistics for bounding box detection. Higher values indicate better results.

| Max Training Boxes | Max Testing Boxes | DSC | IoU | AP IoU=0.50:0.95 | AP IoU=0.50 | AP IoU=0.75 | AR IoU=0.50:0.95 |
|---|---|---|---|---|---|---|---|
| *1* | *1* | 0.54±0.20 | 0.40±0.18 | 0.082 | 0.209 | 0.046 | 0.102 |
| *2* | *2* | 0.62±0.18 | 0.47±0.18 | 0.116 | 0.297 | 0.056 | 0.148 |
| *3* | *3* | 0.60±0.17 | 0.45±0.17 | 0.120 | 0.302 | 0.061 | 0.156 |
| *4* | *4* | 0.57±0.17 | 0.41±0.16 | 0.120 | 0.309 | 0.054 | 0.160 |
| *5* | *5* | 0.58±0.19 | 0.43±0.17 | 0.120 | 0.318 | 0.060 | 0.158 |
| *1* | *2* | 0.62±0.18 | 0.48±0.17 | 0.117 | 0.301 | 0.059 | 0.152 |
| *3* | *2* | 0.62±0.18 | 0.47±0.18 | 0.116 | 0.290 | 0.062 | 0.151 |
| *4* | *2* | 0.62±0.18 | 0.47±0.18 | 0.117 | 0.305 | 0.050 | 0.153 |
| *5* | *2* | 0.64±0.16 | 0.49±0.16 | 0.119 | 0.313 | 0.061 | 0.153 |

**Table 2.** Network performance statistics for carotid plaque segmentation prediction.

| Model | Dice | IoU |
|---|---|---|
| Three-Channel | 0.69±0.21 | 0.56±0.22 |
| Single Channel (B-Mode) | 0.69±0.20 | 0.55±0.20 |

Table 3. Comparison of input data results using the best performing model (3 max training boxes and 2 max testing boxes).

| Model | DSC | IoU | AP IoU= 0.50:0.95 | AP IoU=0.50 | AP IoU=0.75 | AR IoU= 0.50:0.95 |
|---|---|---|---|---|---|---|
| Lumen model | **0.89±0.07** | **0.81±0.11** | **0.62** | **0.98** | **0.68** | **0.68** |
| Best Plaque model | 0.74±0.17 | 0.61±0.20 | 0.159 | 0.334 | 0.140 | 0.205 |
| Filtered Plaque Predictions | 0.76±0.17 | 0.61±0.21 | | | | |

Table 4. Comparison of network performance results for Lumen and plaque models

# Discussion

Of the 9 models trained and tested using different bounding box restrictions, the model trained with a maximum of five bounding boxes and tested on a maximum of two bounding boxes achieved the best performance based on DSC (0.74 for bounding box and 0.64 for segmentation) and IoU(0.61 for bounding box and 0.49 for segmentation). For each model of the Plaque Mask R-CNN presented here, the resulting mean bounding box Dice scores are consistent. The models, when limited to two bounding box predictions in testing, produce the best overall results for each model. However, the results of individual image scores vary amongst the different models as shown in Figure 2. Looking particularly at cases II and III, each of the three models produce differing results. Figure 2(c) shows good agreement for both plaque segments, while 2(f) both include a false positive detection, as well as a false negative detection for the near wall plaque. For the models with 4 or 5 maximum bounding box predictions, many cases can end up with more False Positive detections than a model trained on 1, 2 or 3 bounding boxes as the maximum number of ground truth boxes in the dataset is 3. As most dataset images have 1 or 2 plaque segments, when the models are limited to 2 boxes in testing, this limits the number of false positives seen and thus produces a better mean Dice score. In some instances, the 4 or 5 bounding box predictions can detect plaque segments left out by other models due to lower prediction confidence. This is seen in Figure 2(i) which recovers the near wall plaque lost in 2(f) but unfortunately still retains a false positive detection.

The variation seen between models is likely attributed to the limited variation within the training dataset. If the training is limited to too few bounding boxes, the network may predict a false positive region over a true positive due to less confidence. If the number of bounding boxes is increased, we may get the same false positive region, but it may also correctly detect the desired true positive region. Limiting the testing bounding boxes to 2 keeps the 2 most confident predictions from the trained model. For most some cases such as Figure 2(I) this can reduce the prevalence of false positives but can also cause some false positives to replace true positive detections. As shown in Tables 1 and 2, this increases the mean DSC and IoU slightly compared to using a one-to-one ratio for maximum training and testing bounding boxes. A more comprehensive dataset would likely decrease the impact of this effect.

Table 3 shows a comparison between data input methods described in our previous work for the best of the 9 independent models (3 maximum training bounding boxes, 2 maximum test bounding boxes). Testing was performed for each of the 9 models, but no significant difference between models was observed as shown in Table 3. Thus, using our modified mask R-CNN for carotid plaque detection we do not see a significant accuracy difference between input methods seen previously for carotid lumen detection.

The incorporation of a method to combine the models seen in Figure 3 allows for an accumulation of the best results on an image-by-image basis. With a mean Dice score of 0.76, this allows for a consensus result for the Plaque Mask R-CNN model as these are the most agreed upon results for each model. However, as shown in Figure 3(b and c) there exists cases where too few models agree upon the correct region and/or too many models identify a false positive region. In both cases the model agreement is higher in false positive regions than in some 'true' regions. This does increase overall performance as shown in Table 1, but as a whole these models consistently provide some erroneous results.

We did not include any multi-comparison corrections in this study as all models proved to be significantly independent (all p-values <0.001) from the lumen segmentation model. Although there would be some inherent variability in model results the filtered combined bounding box results show that the models produce mostly similar results with only a small increase in DSC (0.76±0.17) compared to the best individual model (0.74±0.17) as shown in Table 4. This would indicate that the plaque segmentation Mask R-CNN model cannot match the results of the lumen segmentation Mask R-CNN in its current state. This is most likely due to a higher variation of plaque presentation amongst patients compared to the carotid vessel lumen, which is typically more uniform amongst patients. To achieve the same accuracy of the Lumen Mask R-CNN, we would need to incorporate more patients with varying plaque structures or by perhaps restricting the search region to the vessel regions identified by the Lumen Mask R-CNN.

This study is limited by the relatively small and homogeneous dataset (118 patients with severe stenosis), which may not represent the full spectrum of plaque morphology. Manual ground truth segmentations were derived by a single expert observer, which may introduce bias. In addition, the Mask R-CNN model was evaluated on static frames at end diastole; dynamic changes or artifacts present in clinical workflows were not assessed.

## Conclusion

In this study, we adapted a previously developed Mask R-CNN model—originally used for carotid lumen segmentation—for the task of carotid atherosclerotic plaque segmentation using RF-derived B-mode ultrasound images. We evaluated nine model configurations with varying limits on the number of predicted bounding boxes during training and testing. Our results show that the best-performing configuration (trained with 5 boxes, tested with 2) achieved a Dice score of 0.74 and IoU of 0.61 for bounding box detection, while a filtered combination of all model outputs marginally improved the Dice score to 0.76.

Despite this performance, the plaque segmentation model did not reach the accuracy of our previously reported lumen model (Dice = 0.89, IoU = 0.81), highlighting the greater variability in plaque morphology. These results suggest that plaque segmentation is more challenging and sensitive to training data diversity and model configuration.

Limitations of this study include the relatively small dataset, limited plaque morphology diversity, and exclusive use of static end-diastolic frames. Future work will involve expanding the training dataset, incorporating transformer-based architectures for comparison, and integrating lumen-guided ROI constraints to improve plaque localization. Ultimately, this work contributes toward enabling automated plaque segmentation in ultrasound elastography pipelines to support large-scale vascular risk assessment.

## Acknowledgments

This research was supported in part by the National Institutes of Health grant R01-HL147866. Ultrasound RF data on patients was acquired with funding from National Institutes of Health grant R01-NS064034. We are grateful to Siemens Medical Solutions USA, Inc., for providing the S2000 Axius Direct Ultrasound Research Interface (URI) and software licenses.

## Conflict of Interest Statement

The authors declare the following financial interests/personal relationships which may be considered as potential competing interests: Tomy Varghese, Carol C. Mitchell, Robert J. Dempsey, Maxwell J. Kiernan, Rashid Al Mukaddim, Jenna Maybock reports financial support was provided by National Institutes of Health grant R01-HL147866. C. Mitchell: Elsevier, author textbook chapters, and W. L. Gore & Associates contracted research grants to University of Wisconsin-Madison, consulting Acoustic Range Estimates. T. Varghese; Research agreement with Siemens Medical Solutions USA, Inc. for acquisition of radiofrequency data used in this paper. S. Wilbrand: No disclosures

## Data available statement

Data is available upon request. Please send an email to Tomy Varghese (tvarghese@wisc.edu) to inquire about data sharing. Once the request is received IRB agreements have to be reached with the requesting author's affiliated institution and UW-Madison.

# Ethics statement

This study was approved by the UW-Madison institutional review board. Volunteers participated in the study only after providing informed consent using a protocol approved by the University of Wisconsin-Madison institutional review board.